\documentclass[manuscript]{aastex631}

\usepackage{amsmath}
\usepackage{multirow}

\newcommand{\bma}[1]{\boldsymbol{#1}}

\shorttitle{Thermal Continuum}
\shortauthors{Keppens et al.}

\begin{document}
\title{The hydrodynamic thermal continuum, with applications to stratified atmospheres and 1D coronal loop models}
\correspondingauthor{Keppens Rony}
\email{keppens.rony@kuleuven.be}

\author[0000-0003-3544-2733]{Rony Keppens}
\affiliation{Centre for mathematical Plasma-Astrophysics (CmPA), KU Leuven \\
Celestijnenlaan 200B, 3001 Leuven, Belgium}

\author[0000-0003-2443-3903]{Jordi De Jonghe}
\affiliation{Centre for mathematical Plasma-Astrophysics (CmPA), KU Leuven \\
Celestijnenlaan 200B, 3001 Leuven, Belgium}

\author[0009-0006-8524-008X]{Adrian Kelly}
\affiliation{Centre for mathematical Plasma-Astrophysics (CmPA), KU Leuven \\
Celestijnenlaan 200B, 3001 Leuven, Belgium}

\author[0000-0002-7885-4554]{Nicolas Brughmans}
\affiliation{Centre for mathematical Plasma-Astrophysics (CmPA), KU Leuven \\
Celestijnenlaan 200B, 3001 Leuven, Belgium}

\author[0000-0002-8794-472X]{Hans Goedbloed}
\affiliation{DIFFER - Dutch Institute for Fundamental Energy Research, De Zaale 20, 5612 AJ Eindhoven, the Netherlands}

\begin{abstract}
 Using both analytical and numerical means, we demonstrate that linear stability analysis of a hydrodynamic stratified atmosphere or a 1D coronal loop model in non-adiabatic settings features a thermal continuum corresponding to highly localized eigenfunctions. This thermal continuum can be precomputed, involving the net heat-loss function and its partial derivatives, and is the generalization of the thermal instability introduced by~\citet{Parker1953}. We account for a thermal imbalance, directly affecting thermal instability growthrates.
 We present completely general equations that govern all eigenmodes, including non-adiabatically affected p- and g-modes of the stratified settings. We intend to clarify how linear thermal instability is relevant for solar loops that show spontaneous in-situ condensations, and eliminate recent confusion on specific isochoric routes to linear instability alongside other thermal instability channels. The thermal continuum, previously identified as a crucial ingredient in magnetohydrodynamic eigenmode spectra for coronal loops and atmospheres, drives multithermal aspects across our universe, such as forming solar coronal rain and prominences, or cold cloud creation in intracluster to interstellar medium environments.
\end{abstract}


\section{Introduction}\label{intro}

\subsection{Motivation}
The solar corona is structured in myriads of magnetic loops and contains rarified plasma at million degrees Kelvin. However, observations reveal how spontaneous in-situ condensations show up nearly everywhere. In the strong coronal magnetic field, these cold (10000 K) and dense blobs are then guided to rain down curved magnetic loops, such that an essentially hydrodynamic evolution along a 1D loop often suffices to explain the physics. These coronal rain condensations are manifestations of an energy exchange between the plasma and the radiation field, where optically thin radiative losses preferentially act as a sink for internal energy. This makes the loop plasma liable to a runaway linear thermal instability, originally postulated by \citet{Parker1953}. Observations, theory and numerical simulations meanwhile confirm the essential role played by thermal instability to instigate coronal rain or prominence formation \citep{li2022,donn2024}, and the very same physical ingredients return in model efforts on stellar prominences \citep{Simon2024}, galactic prominences \citep{Peng2017}, cold gas precipitation and survival in galactic stratified settings \citep{Gronke2023,Wibking2025}, or in stellar or galactic outflows where the wind demonstrates multi-thermal structure \citep{Wareing2017}, as influenced or triggered by radiative effects. In this paper, we show that a rigorous mathematical analysis of the linearized non-adiabatic hydrodynamic equations reveals the presence of a thermal continuum in the essential spectrum of eigenmodes of a stratified atmosphere or coronal loop. We will show how p- and g-modes get modified by radiative losses, and that the usually uninteresting (at zero frequency) entropy mode turns into a continuous range of eigenfrequencies with ultra-localized (singular) eigenfunctions. This insight connects and clarifies findings on thermal non-equilibrium states, waves affected by thermal misbalance, and the present confusion on the role of isochoric versus isobaric thermal instabilities~\citep{Waters2025}.

\subsection{Spectroscopy in ideal (magneto)hydrodynamic settings}

In ideal magnetohydrodynamics (MHD), categorizing all linear waves and instabilities accessible to a particular equilibrium configuration is known as magnetohydrodynamic spectroscopy \citep{goed2019}. This spectrum is quantified by normal mode eigenfrequencies $\omega$ arising from assuming an exponential $\exp(-i\omega t)$ temporal variation for all linear quantities.
Since ideal MHD is time-reversible, the entire spectrum located throughout the complex $\omega$-plane has an up-down symmetry: for every overstable or purely unstable mode $\omega=\omega_R+i\omega_I$ (with $\omega_I>0$), there is a corresponding damped counterpart $\omega_R-i\omega_I$ that is also part of the spectrum. In equilibrium configurations that obey an essentially one-dimensional force balance, such as in a gravitationally stratified atmosphere with height variation only, a key role in MHD spectroscopy is played by real-valued (hence, stable) continuous ranges in the spectrum, which locate on the $\omega_R$-axis. In MHD settings, these form pairs of backward and forward Doppler-shifted slow and Alfv\'en continua, so we have four continua that may  overlap. Since MHD counts seven wave families, its essential spectrum is further aware of the infinite (real) frequency $\omega\rightarrow\pm \infty$ accumulation points for fast modes, and a separate Eulerian entropy continuum that locates throughout the range of Doppler shifts. In the ideal hydrodynamic limit, where we expect five wave families, the Eulerian entropy continuum remains, and we find two overlapping flow continua at the same Doppler range, related to singular Eulerian velocity perturbations as discovered by \citet{Case1960}, augmented with the infinite (real) frequency accumulation points for sound waves. When no background flow is considered in the state that is spectrally analyzed, all three overlapping hydrodynamic continua collapse to marginal frequency $\omega=0$, since Doppler shifts vanish. In what follows, we will assume zero flow in the background state, but show how the marginal $\omega=0$ entropy mode becomes a crucial (and continuous) ingredient of the hydrodynamic spectrum, when non-adiabatic effects are included. 

\subsection{The thermal continuum in magnetohydrodynamics}

When non-ideal effects, such as resistivity, thermal conduction, or viscosity, are incorporated, we lose time-reversibility in (M)HD, such that the up-down symmetry in the complex eigenfrequency plane gets broken. 
In this paper, we focus attention to a particular non-adiabatic effect, namely we allow a deviation of entropy-conserving evolutions. Specifically, when a source or sink term for internal energy that represents the (assumed instant and local) optically thin radiative losses is incorporated, \citet{vdl1991} showed that in a cylindrical, static MHD equilibrium configuration, the governing ordinary differential equations for normal modes contain a thermal continuum. This continuum can be purely unstable (i.e. only have nonvanishing $\omega_I>0$) and is coupled to both MHD slow continua, which are also affected by non-adiabatic processes. This thermal continuum is in fact the non-adiabatic counterpart of the Eulerian entropy continuum, and directly generalizes the thermal instability that was postulated to exist in uniform, but radiating plasma by \citet{Parker1953}. Indeed, under typical solar coronal conditions, the radiative losses act destabilizing, with a potential for runaway condensations to form as a direct result. Cylindrical flux ropes were analyzed in \cite{vdl1991B} and slab configurations were treated in \cite{vdlslab1991}. The analysis accounted for anisotropic thermal conduction, with both field-aligned (parallel) heat conduction as well as finite perpendicular thermal conduction $\kappa^c_\perp$, across the magnetic surfaces. Using numerically computed spectra, it was demonstrated that finite perpendicular thermal conduction replaces the thermal continuum by a dense set of discrete thermal modes.

\subsection{Goal and organization of this paper}

The fate of especially this thermal continuum, is as yet undocumented for purely hydrodynamical settings. We do so in this paper, where we recover some of the MHD findings by simply inspecting their zero magnetic field limit. We will at the same time incorporate gravitational stratification, which in the case of stellar seismology brings in the familiar p- and g-modes, for pressure and gravity driven oscillations. It is our goal to present the most general equations governing a gravitationally stratified, non-adiabatically evolving gas or plasma, and perform the manipulations on the linearized equations that show the occurrence of the thermal continuum. 

In Section~\ref{nonlineq}, we present the governing non-linear equations in their most general form, and in various fully equivalent formulations. In Section~\ref{lineareq} we linearize the equations about a static, stratified setting, allowing for a non-perfect balance in the thermodynamics (a thermal misbalance or non-perfect coupling between the radiation field and the plasma). We then apply it to both plane-parallel, gravitationally stratified atmospheres as well as to the frequently adopted one-dimensional, area-expanding hydrodynamic coronal loop models \citep{Mikic2013}. The analytical formulations are presented along with numerical demonstrations in both settings as realized by the open-source eigenvalue solver {\texttt{Legolas}} \citep{legolasA2020,legolasB}. We end with a brief discussion and outlook, emphasizing relations to recent findings. 

\section{Governing nonlinear equations in equivalent forms}
\label{nonlineq}

Using density $\rho$, velocity vector $\bma{v}$, and pressure $p$, the gas dynamical governing equations for mass conservation, momentum (Newton's law) and energy evolution can be written as
\begin{eqnarray}
    \partial_t \rho+\nabla \cdot  \left (  \rho \bma{v}\right ) &=& 0\,,
    \label{dens}\\
    \partial_t \bma{v}+\bma{v}\cdot\nabla \bma{v} +\frac{\nabla p}{\rho} &=& \bma{g}\,,
    \label{mom}\\
    \frac{p}{\gamma-1}\frac{d \left(\ln{S}\right)}{dt} &=& \dot{q} \equiv \rho{\cal{L}} \,.
    \label{entr}
\end{eqnarray}
The latter equation uses the Lagrangian derivative $\frac{d}{dt}\equiv \partial_t +\bma{v}\cdot\nabla$ to quantify how the entropy $S$ changes due to heat exchange, in accord with the first law of thermodynamics, and features the constant ratio of specific heats $\gamma$.

The right-hand-side source terms in the momentum equation~(\ref{mom}) and entropy equation~(\ref{entr}) represent, in general, a combination of physical mechanisms. The acceleration vector $\bma{g}$ will in our applications represent a spatially varying external gravity. For the applications to 1D coronal loop models, the shape of the fixed magnetic field determines the spatial dependence of $\bma{g}$, which is then effective only along the field line or coronal ``loop". The source term $\dot{q}$ in the entropy equation~(\ref{entr}) represents the collection of all non-adiabatic effects one wishes to consider for the gas/plasma evolution. It could include thermal conduction (we will discuss this effect separately in Section~\ref{tcpart}), but we focus our discussion on the internal energy exchange between the gas or plasma and a background radiation field. The term $\dot{q}$ indicates the net heating and cooling per unit volume, while the equivalent notation $\rho{\cal{L}}$ rather focuses on the net heating-cooling per unit mass.

We will adopt an ideal gas law with a fixed gas constant ${\cal{R}}=k_B/(\mu m_p)$, involving Boltzmann constant $k_B$, proton mass $m_p$ and fixed mean molecular weight $\mu$, such that we here ignore composition/ionization effects. The ideal gas law writes generally as follows, and we recall how the ratio of specific heats $\gamma$ is related to the specific heats at constant volume and pressure:
\begin{eqnarray}
     p & = & {\cal{R}}\rho T \,, \label{idealgas} \\
    C_v &= & \frac{{\cal{R}}}{\gamma-1}\,, \,\,\,\,\,\,\,\,\,
    C_p  =  \gamma C_v = C_v+{\cal{R}}= \frac{\gamma {\cal{R}}}{\gamma-1}\,. \label{specheats}
\end{eqnarray}
The entropy $S$ can also be expressed directly in terms of the thermodynamic variables $\rho$ and $p$, and the first law of thermodynamics in Eq.~(\ref{entr})
can be written using a specific entropy per unit mass $S^*$ 
\begin{eqnarray}
T\frac{d S^*}{dt} & = & {\cal{L}} \,, \label{entrspec}
\end{eqnarray}
which is related to the dimensionless $S$ and density-pressure by
\begin{eqnarray}
S^* &=& C_v \ln{S}= C_v \ln{\left[\frac{p}{p_{u}}\left(\frac{\rho}{\rho_u}\right)^{-\gamma}\right]} \,. \label{entropy}
\end{eqnarray}
In this writing, we introduced reference units for pressure $p_u$ and density $\rho_u$, and we could similarly measure temperature in $T_u$ units such that $p_u={\cal{R}}\rho_uT_u$. In terms of $S^*$ the first law of thermodynamics can be written in a number of equivalent formulations, where we only adopt the ideal gas law from Eq.~(\ref{idealgas}), to relate the Lagrangian changes in $S^*$ with those in thermodynamic variables through
\begin{eqnarray}
      \frac{dS^*}{dt} &=& C_v\frac{1}{p}\frac{dp}{dt}-C_p \frac{1}{\rho}\frac{d\rho}{dt}  = C_p\frac{1}{T}\frac{dT}{dt}-{\cal{R}} \frac{1}{p}\frac{dp}{dt} = C_v\frac{1}{T}\frac{dT}{dt}-{\cal{R}} \frac{1}{\rho}\frac{d\rho}{dt} =\frac{\cal{L}}{T}\,. \label{enC}
\end{eqnarray}
In practice, this implies that instead of the entropy formulation Eq.~(\ref{entr}), we could also work with the completely equivalent pressure or temperature evolution equations formulated as
\begin{eqnarray}
    \partial_t p+\bma{v}\cdot \nabla p + \gamma p \nabla\cdot\bma{v} & = & \left(\gamma-1\right)\dot{q} \,,\label{pevol}\\
    {\cal{R}}\rho \partial_t T +{\cal{R}}\rho\bma{v}\cdot\nabla T + (\gamma-1)p\nabla\cdot \bma{v} &=& \left(\gamma-1\right)\dot{q} \label{Tevol}\,.
\end{eqnarray}
In what follows, we will provide fully redundant expressions for all these formulations, as it will be useful to discuss specific linear evolutions that keep a certain thermodynamic variable constant. Now an important first observation is the following: whether one chooses the $\rho,\bma{v},S$, or the $\rho,\bma{v},p$ or the $\rho,\bma{v},T$ formulation, there are always exactly five partial derivatives $\partial_t$ with respect to time, implying that in a uniform medium where we can Fourier analyze in all directions, we expect exactly five wave modes to occur. In what follows, we perform the linear analysis of these governing non-adiabatic equations.

\section{Linearized equations: general formulations}
\label{lineareq}

The nonlinear equations from above can be linearized about any state (even one drawn from a fully nonlinear, time-dependent evolution as demonstrated in \cite{Keppens2016}), but it is customary to adopt a background that at least obeys some or all of the governing equations at zeroth order. In this work, we will consider a background that obeys mass conservation and force balance exactly, but does not have a perfect balance in the energy equation. We take a static $\bma{v}_0=0$, stratified medium, with a thermal misbalance expressed as
\begin{eqnarray}
\bma{v}_0 & =& \bma{0} \,, \,\,\,\,\, 
    \nabla p_0  = \rho_0 \bma{g}(\bma{x}) \,, \,\,\,\,
    {{\cal{L}}_0}  \neq  0 \,. \label{equist}
\end{eqnarray}
The thermal imbalance implies that we do not really have a zeroth order, time-independent solution to the energy balance, as in terms of background temperature $T_0$ we find from Eq.~(\ref{Tevol}) that
\begin{equation}
    C_v\partial_t T_0={\cal{L}}_0 \,. \label{T0evol}
\end{equation}
Nevertheless, we can still compute the entire normal mode spectrum for the background state ignoring its time-variation, which we will do in what follows. That we do not strictly obey all zeroth order equations is similar to the common practice of computing resistive MHD spectra for purely force-balanced ideal MHD configurations, such as the well-known Harris sheet equilibrium for quantifying tearing modes: in that setting the resistive diffusion of the current sheet is assumed to happen much slower than the growth rate found for tearing modes. Since Eq.~(\ref{T0evol}) introduces its own (local) timescale $\tau_0=C_v T_0/{\cal{L}}_0$ for adjusting the background state, we encounter the possibility for time-scale conflicts between growthrates as found in the instantaneous eigenspectrum, and the equilibrium evolution. When ${\cal{L}}_0=0$ everywhere, $\tau_0$ becomes infinite as we then analyze an exact zeroth order, time-independent solution to all governing equations.

When we linearize about the state~(\ref{equist}), we will encounter terms we can express as entropy gradients, computed from
\begin{equation}
    \frac{\nabla S_0}{S_0} = \frac{\nabla p_0}{p_0}-\gamma \frac{\nabla \rho_0}{\rho_0} = \frac{1}{p_0}\left[\rho_0\bma{g}-c_0^2\nabla\rho_0\right] \,,
\end{equation}
where we introduced the squared sound speed
$c_0^2=\gamma{\cal{R}}T_0=\gamma p_0/\rho_0$. Note that this leaves the possibility to have unaligned pressure, density and entropy gradients, and it contains the special case of a constant entropy background ($S_0=constant$). For the background from Eq.~(\ref{equist}), we find as linearized equations
\begin{eqnarray}
    \partial_t \rho_1 + \nabla \cdot\left(\rho_0\bma{v}_1\right) & = & 0\,, \label{linrho}\\
    \rho_0\partial_t \bma{v}_1+\nabla p_1 - {\rho_1}\bma{g} & = & \bma{0} \,, \label{linmom}\\
    \frac{p_0}{(\gamma-1)S_0}\left[\partial_t S_1 +\bma{v}_1\cdot\nabla S_0\right] & = & \dot{q}_1 \equiv \rho_1{{\cal{L}}_0}+\rho_0{{\cal{L}}_1}= \rho_1{{\cal{L}}_0}+\rho_0\left\{\left[{\cal{L}}_{\rho|S}\right]\rho_1+\left[{\cal{L}}_{S|\rho}\right]S_1\right\}\,,\label{linen}
\end{eqnarray}
where the last Eq.~(\ref{linen}) introduced the linearized version of the net heat/loss term. Here we made the assumption that ${\cal{L}}$ depends on only two thermodynamic variables, and in such a case, there are clear relations between the partial derivatives of ${\cal{L}}$ to thermodynamic variable $A$ when keeping variable $B$ constant, which we denoted in the last equation as ${\cal{L}}_{A|B}$. We will denote ${\cal{L}}_{\rho|T}\equiv {\cal{L}}_\rho$ and ${\cal{L}}_{T|\rho}\equiv {\cal{L}}_T$ and express all others in function of these two.
Under the ideal gas law, we find
\begin{eqnarray}
    {\cal{L}}_1 & = & {\cal{L}}_{\rho}\rho_1+{{\cal{L}}_T}T_1 \,,\\
     &=& \left({\cal{L}}_{\rho}-\frac{T_0}{\rho_0}{\cal{L}}_T\right)\rho_1 +\frac{{\cal{L}}_T}{{\cal{R}}\rho_0}p_1 \,, \\
     &=& \left({\cal{L}}_{\rho}-\frac{T_0}{\rho_0}(1-\gamma){\cal{L}}_T\right)\rho_1 +\frac{T_0{\cal{L}}_T}{S_0}S_1 \,.
\end{eqnarray}
We can easily manipulate the 5 linearized equations from Eq.~(\ref{linrho}-\ref{linen}) to 
\begin{equation}
    \nabla\left[c_0^2\nabla \cdot\left(\rho_0\bma{v}_1\right)-\frac{p_0}{S_0}\partial_t S_1\right] -\bma{g}\nabla\cdot\left(\rho_0\bma{v}_1\right) -\rho_0\partial^2_{tt} \bma{v}_1 = \bma{0} \,. \label{genvS}
\end{equation}
Note that this equation still contains the Eulerian time derivative $\partial_t S_1$, which for vanishing ${\cal{L}}$, would write directly as $-\bma{v}_1\cdot \nabla S_0$. 

Let us first consider the possibility for exactly isobaric solutions where $p_1=0$. Then, directly from only the first two Eqns.~(\ref{linrho}-\ref{linmom}) we get 
\begin{equation}
     -\bma{g}\nabla\cdot\left(\rho_0\bma{v}_1\right) =\rho_0\partial^2_{tt} \bma{v}_1 \,,
\end{equation}
while the remaining Eq.~(\ref{linen}) then demands, if we also Fourier analyze in time so adopt $\exp{(-i\omega t)}$, that
\begin{equation}
    -i M\rho_1=p_0 \bma{v}_1\cdot \frac{\nabla S_0}{S_0} \,,
\end{equation}
where we introduce
\begin{equation}
    M   = \omega c_0^2 + i (\gamma-1) \left({{\cal{L}}_0}+\rho_0{\cal{L}_{\rho}}-T_0{\cal{L}}_T\right) \,.
\end{equation}
This expression will lead to the thermal continuum.

Generally (i.e. not assuming isobaric conditions), a Fourier treatment for time allows to derive the following relations
\begin{eqnarray}
    i\omega \rho_1 & = & \nabla \cdot\left(\rho_0 \bma{v}_1\right) \,, \label{idrho}\\
    i D \frac{S_1}{S_0} & = & \bma{v}_1 \cdot \frac{\nabla S_0}{S_0} - \frac{Q}{p_0}\frac{\rho_1}{\rho_0} =iD\left[\frac{p_1}{p_0}-\gamma\frac{\rho_1}{\rho_0}\right]\,, \label{idS} \\
     i D \frac{p_1}{p_0} & = & \bma{v}_1 \cdot \frac{\nabla S_0}{S_0} + \frac{i M}{{\cal{R}}T_0}\frac{\rho_1}{\rho_0} =iD\left[\frac{T_1}{T_0}+\frac{\rho_1}{\rho_0}\right]\,, \label{idp} \\
      i D \frac{T_1}{T_0} & = & \bma{v}_1 \cdot \frac{\nabla S_0}{S_0} + \left[iD(\gamma-1)-\frac{Q}{p_0}\right]\frac{\rho_1}{\rho_0} = iD \left[\frac{S_1}{S_0}+(\gamma-1)\frac{\rho_1}{\rho_0}\right]\,. \label{idT}
\end{eqnarray}
The last three of these relations identify how entropy, pressure or temperature perturbations relate to the velocity perturbation and the perturbed density $\rho_1$, while the rightmost equalities use the ideal gas law and the definition of entropy. Since Eq.~(\ref{idrho}) relates density to velocity, it is clear that we can express all linearized equations directly in terms of the velocity $\bma{v}_1$. The spatially dependent coefficients $D$ and $Q$ in these relations~({\ref{idS}-\ref{idT})} are given by
\begin{eqnarray}
    D & = & \omega-i(\gamma-1)\frac{{{\cal{L}}_T}}{\cal{R}}=\omega-i\frac{{\cal{L}}_T}{C_v} \,, \\
    Q & = & (\gamma-1) \rho_0 \left({{\cal{L}}_0}+\rho_0{\cal{L}}_\rho-T_0(1-\gamma){\cal{L}}_T\right) \,, 
    \end{eqnarray}
    while we note their relation to $M$ as
    \begin{equation}
    M   =  c_0^2 D +\frac{iQ}{\rho_0}\,.
\end{equation}
Similarly, we can find a general expression linking temperature perturbations $T_1$ to velocity and pressure, namely
\begin{eqnarray}
      \left(i D\gamma -\frac{Q}{p_0}\right) \frac{T_1}{T_0} & = & \bma{v}_1 \cdot \frac{\nabla S_0}{S_0} + \left[iD(\gamma-1)-\frac{Q}{p_0}\right]\frac{p_1}{p_0} \,. \label{idTB} 
\end{eqnarray}

Using any of the $\rho,\bma{v},S$, or the $\rho,\bma{v},p$ or the $\rho,\bma{v},T$ formulation, we now find easily that we can reduce the system of 5 equations to a single vector equation given by
\begin{equation}
    \nabla \left[\frac{M}{D}\nabla\cdot\left(\rho_0\bma{v}_1\right)+\frac{\omega}{D}\bma{v}_1\cdot\left(\rho_0\bma{g}-c_0^2\nabla\rho_0\right)\right]-\bma{g}\nabla\cdot\left(\rho_0\bma{v}_1\right)+\rho_0\omega^2 \bma{v}_1 = \bma{0} \,. \label{generalv1}
\end{equation}
This equation governs all waves and instabilities in any medium obeying Eq.~(\ref{equist}). We will now manipulate this equation further for the specific cases of plane-parallel stratified media (Section~\ref{sec:atm}), or 1D coronal loops (Section~\ref{sec:loop}). We first make the link to known formulations in adiabatic, nonuniform settings (Section~\ref{sec:ad}) or non-adiabatic findings for uniform settings (Section~\ref{sec:uni}).

Before doing so, we note that there seems to be a `trivial' solution, namely one without flow perturbations $\bma{v}_1=\bma{0}$, and with $\rho_1=0$ (i.e. static and isochoric), if we take $D=0$, leaving $S_1,p_1,T_1$ arbitrary in accord with Eqns.~(\ref{idrho})-(\ref{idT}). This is the observation made by \citet{Waters2025}, calling it the isochoric Catastrophic Cooling or CC mode. However, this is not a generally valid eigenmode solution, since we must realize that $D(\bma{x})$ is spatially dependent, and whatever net heating-cooling functionality is adopted for ${\cal{L}}$ (see further in Section~\ref{sec:lform}), its partial derivative ${\cal{L}}_T$ evaluated for the equilibrium will almost surely not be spatially uniform. Hence, you can only have a local $D(\bma{x}_*)=0$, and correspondingly allow delta-type solutions with all other thermodynamic variables $T_1(\bma{x}_*), p_1(\bma{x}_*),S_1(\bma{x}_*)\neq 0$, but vanishing everywhere else. In \citet{Waters2025}, a spatially homogeneous $T_1(t)$ was argued instead, which is clearly impossible in general. The fact that delta-type solutions seem an option throughout the range $D=0$ suggests that this range may represent a continuum of purely unstable or damped modes, that can be computed directly: their temporal growth/decay would be given by $\exp{(t{\cal{L}}_T/C_v)}$ and a positive ${\cal{L}}_T$ implies instability. This possibility was first pointed out in \citet{Parker1953}, where only the temperature evolution equation in the absence of flow (i.e. equation~(\ref{Tevol}) for $\bma{v}=\bma{0}$) was perturbed and discussed (including thermal conduction). However, we will demonstrate that instead, the factor $M=0$ corresponds to a genuine continuum of thermal instabilities, and that the continuous range $D=0$ is only signaling apparent singularity, in complete analogy with the occurrence of genuine and apparent continuous singularity ranges, identified in rigorous linear spectral theory for ideal magnetohydrodynamics \citep{goed2019,Hans1998,Appert1974}. It is therefore useful to recall the purely adiabatic case.

\subsection{Adiabatic reduction}\label{sec:ad}
When ${\cal{L}}=0$, we find that
$D^{ad}=\omega$, $Q^{ad}=0$, $M^{ad}=c_0^2\omega$ so Eq.~(\ref{generalv1}) reduces to
\begin{equation}
    \nabla \left[c_0^2\nabla\cdot\left(\rho_0\bma{v}_1\right)+{p_0}\bma{v}_1\cdot \frac{\nabla S_0}{S_0}\right]-\bma{g}\nabla\cdot\left(\rho_0\bma{v}_1\right)+\rho_0\omega^2 \bma{v}_1 = \bma{0} \,. \label{adreduc}
\end{equation}
In this adiabatic case, textbook formulations \citep{goed2019} employ a Lagrangian displacement vector $\bma{\xi}$, which is linked to $\bma{v}_1=\partial_t\bma{\xi}=-i\omega \bma{\xi}$ due to the assumption of a static background where $\bma{v}_0=\bma{0}$.
This Lagrangian viewpoint can exploit the self-adjoint force operator $\bma{F}(\bma{\xi})$ which writes Eq.~(\ref{adreduc}) for the hydro case rather as
\begin{equation}
    \bma{F}(\bma{\xi})\equiv \nabla\left(\gamma p_0\nabla\cdot \bma{\xi}+\bma{\xi}\cdot\nabla p_0\right)-\bma{g}\nabla\cdot(\rho_0 \bma{\xi})=\rho_0\partial^2_{tt} \bma{\xi} \,.
    \label{adiabcase}
\end{equation}
Note that we here reverted to general time-dependent $\bma{\xi}(\bma{x},t)$ as directly obtained from Eq.~(\ref{genvS}).

This switch to a Lagrangian description allows focus on non-trivial waves (essentially sound waves, modified by gravity) only, since Eq.~(\ref{linrho}) can be integrated to $\rho_1=-\nabla\cdot (\rho_0 \bma{\xi})$, while the entropy equation~(\ref{linen}) now merely states that the Lagrangian entropy perturbation given by $S_{1L}=S_1+\bma{\xi}\cdot\nabla S_0$ vanishes. The self-adjoint force operator formalism shows us that only real $\omega^2$ are possible for adiabatic conditions, so that either stable waves ($\omega^2>0$) or purely unstable or damped ($\omega^2<0$) solutions exist. In the complex eigenfrequency plane, this means that modes locate either on the $\omega_R$ or the $\omega_I$ axis, keeping left-right and up-down symmetry.

Specifying to a uniform medium leaves only the first and last term in Eq.~(\ref{adreduc}) or Eq.~(\ref{adiabcase}), giving forward and backward sound waves $\omega^2=k^2c_0^2$ when introducing a wavevector $\bma{k}$ with wavenumber $k$. The other three modes are all at marginal frequency $\omega=0$, but they come in two flavors: one corresponds to the entropy mode where $\bma{v}_1=\bma{0}$ and $p_1=0$, but $\rho_1,S_1,T_1$ arbitrary: this is the one related to the thermal instability we will discuss shortly for non-adiabatic regimes. The other two marginal solutions are shear waves, which become simple transverse translations where $\bma{v}_1\perp\bma{k}$.

\subsection{Non-adiabatic generalization for a  uniform medium}\label{sec:uni}

As soon as we consider non-adiabatic effects, the eigenfrequency $\omega$ returns in factors $D$ and $M$ in our Eq.~(\ref{generalv1}), so unlike the adiabatic case, we no longer have a self-adjoint formulation. We will first discuss the uniform medium in non-adiabatic settings. Of course, a uniform medium implies $\nabla S_0=\bma{0}$ and can not consider gravity $\bma{g}=\bma{0}$, but allows to introduce the wavevector $\bma{k}$ and we find from Eq.~(\ref{idrho}) and (\ref{idS})-(\ref{idp}) that
\begin{eqnarray}
    \omega \rho_1 & = & \rho_0\bma{k} \cdot\bma{v}_1 \,,\\  
    \frac{S_1}{S_0}& =& \left(\frac{M\rho_0}{Dp_0}-\gamma\right)\frac{\rho_1}{\rho_0}\,,
\end{eqnarray}
while the dispersion relation is obtained directly from Eq.~(\ref{generalv1}) as
\begin{eqnarray}
 -\frac{M}{D}\bma{k}\left(\bma{k}\cdot\bma{v_1}\right)+\omega^2\bma{v}_1 &= &\bma{0} \,.
 \end{eqnarray}
This dispersion relation leaves two possibilities, namely 
\begin{equation}
    (1) \,\,\, \omega=0 \,\,\,\mathrm{and}\,\,\,\bma{k}\cdot\bma{v}_1=0 \,\,\,\,\,\,\mathrm{or}\,\,\,\,\,(2)\,\,\, \omega^2 D-k^2M=0 \,. \label{criteria}
\end{equation}
The last equation writes out in full as a third order polynomial in $\omega$, given by
\begin{equation}
    \omega^3-i\omega^2\frac{{\cal{L}}_T}{C_v}-\omega c_0^2 k^2-i(\gamma-1)\left({\cal{L}}_0+\rho_0{\cal{L}}_\rho-T_0{\cal{L}}_T\right)k^2=0 \,. \label{druni}
\end{equation}
We have recovered known results here: there are
two trivial shear waves with $\bma{k}\cdot\bma{v}_1=0$ at $\omega=0$, which do not feature in the dispersion relation Eq.~(\ref{druni}). The relation (\ref{druni}) combines the entropy mode with the (forward and backward) sound waves. This cubic polynomial in $\omega$ can be rewritten to one with only real coefficients, by replacing $-i\omega\equiv n$, and any such cubic polynomial has either three real, or one real and two complex conjugate solutions. This was discussed at length in \citet{Field1965} (where it was extended to MHD as well) and the hydro regime was revisited in \citet{Waters2019}. The two complex conjugate sound wave solutions can be overstable or damped, or even become purely growing/damped solutions. The $\omega=0$ marginal entropy mode from the adiabatic uniform medium becomes the thermal instability. In \citet{Waters2019}, so-called `fast' and `slow' isochoric condensation modes were introduced when the `sound waves' turn into purely real $n$, quantifying the critical wavelengths where the two complex conjugate solutions would become real-valued. In the context of more general MHD spectral theory, calling the corresponding purely growing `slow' and `fast' condensation modes may introduce confusion, since the hydro limit has no true slow or fast magneto-acoustic modes. Generalizations of dispersion relation~(\ref{druni}) are known for cases modified by thermal conduction (see Section~\ref{tcpart}), where background lengthscales like the Field length \citep{Begelman1990} would enter, to be compared with the wavelength $\lambda=2\pi/k$ corresponding to the wavenumber $k$. Isentropic, isobaric, or isochoric behavior can be studied and related to the so-called S-curve, which is found from a perfect balance in heating and cooling or ${\cal{L}}=0$, and is usually drawn in a state-space view, hence as a relevant curve in e.g. the $(T,1/\rho)$-plane \citep{Waters2019,herm2021}. 

Returning to the isochoric instability as advocated by \citet{Waters2025}, it is true that Eq.~(\ref{criteria}) or Eq.~(\ref{druni}) has a formal solution for $k=0$ (i.e. a spatially homogeneous mode) that corresponds to $D=0$. However, this is specific to truly homogeneous media extending infinitely in all directions, a situation that does not occur in practice. It would also then represent a linear perturbation requiring infinite energy.

Extensions of the dispersion relation Eq.~(\ref{druni}) to non-adiabatic, still uniform, but truly magnetohydrodynamic settings, were presented in \citet{Field1965}, but we recently revisited these to highlight the eigenfunction polarizations, and how these relate to condensation formation in local box simulations of the solar corona \citep{Claes2019,Claes2020,herm2021}. We note that the statement we made in \citet{Claes2019}, on how isochoric versus isobaric growth rates can be found by multiplying selected terms in the dispersion relation Eq.~(\ref{druni}) is incorrect: there are no other modes than those described by the governing Eqns.~(\ref{criteria})-(\ref{druni}). 

\subsection{Non-adiabatic generalization with 1D equilibrium variation}
\label{sec:atm}

Here, we analyze the non-adiabatic case in the presence of 1D inhomogeneity.
We will then additionally assume that $\nabla \rho_0$ is parallel to the already parallel directions $\bma{g}$ and $\nabla p_0$. Then, under the ideal gas law, also $\nabla T_0$ is in that direction, as well as $\nabla S_0$. This implies we can call this physically meaningful direction $x$, making $p_0(x),\rho_0(x), T_0(x), S_0(x)$ and denote gravity as $\bma{g}=g_{\parallel}(x)\hat{\bma{e}}_x=-g(x)\hat{\bma{e}}_x$. We will then also assume the same purely 1D behavior for ${\cal{L}}$, so ${\cal{L}}_0(x)$, as typically through ${\cal{L}}(\rho,T)$. Derivatives on the configuration quantities are then denoted by $d\rho_0/dx\equiv \rho_0'$.

\subsubsection{General formulations}
The 1D variation of the background state implies the possibility to Fourier analyze the ignorable directions $y$ and $z$, so we can write for every linearized velocity component $v_{1x}=\hat{v}_{1x}(x)\exp{(ik_yy+ik_zz})$. Our general equation~(\ref{generalv1}) then rewrites as
\begin{equation}
    \left(\begin{array}{ccc}
         \frac{d}{dx}\left(\frac{N}{D}\frac{d}{dx}+\frac{W}{D}\right)+\rho_0{\tiny(\omega^2-g')} \,\,\, & ik_y\left(\frac{d\,}{dx} \frac{N}{D}+\rho_0 g\right) \,\,\,& ik_z\left(\frac{d}{dx}\frac{N}{D} +\rho_0 g\right) \\
         ik_y\left(\frac{N}{D}\frac{d\,}{dx} -\rho_0 g+\frac{W}{D}\right)& -\frac{N}{D}k_y^2+\rho_0\omega^2 & -\frac{N}{D}k_yk_z  \\
         ik_z\left(\frac{N}{D}\frac{d\,}{dx} -\rho_0 g+\frac{W}{D}\right)& -\frac{N}{D}k_y k_z &  -\frac{N}{D}k_z^2+\rho_0\omega^2
    \end{array} \right)
    \left(\begin{array}{c}
         \hat{v}_{1x}  \\
          \hat{v}_{1y}  \\
           \hat{v}_{1z}  \\
    \end{array}\right)
    = 
    \left(\begin{array}{c}
         0 \\
         0 \\
         0
    \end{array}\right)\,, \label{matform} \end{equation}
where we introduced
\begin{eqnarray}
    N & =& M\rho_0 \,,\\
    W & = & i(\gamma-1) \left[\rho_0'\left({\cal{L}}_0+\rho_0{\cal{L}}_\rho-T_0{\cal{L}}_T\right)-\rho_0 g\frac{{\cal{L}}_T}{\cal{R}}\right] \,,\\
      & = & \rho_0'\left(M-\omega c_0^2\right)-p_0'(D-\omega) \,.
\end{eqnarray}
Due to the invariance in the $y-z$ plane, we can
align to the wavevector by introducing $k_0=\sqrt{k_y^2+k_z^2}$ without loss of generality, and find from only the first two equations~(\ref{matform}) that
\begin{equation}
  \hat{v}_{1y}=-ik_0 \frac{\frac{N}{D}\frac{d\hat{v}_{1x}}{dx}+\left(\frac{W}{D}-\rho_0g\right)\hat{v}_{1x}}{\rho_0\omega^2-k_0^2\frac{N}{D}}  \,,
\end{equation}
while the relevant governing second order ODE becomes
\begin{multline}
    \frac{d}{dx}\left(\frac{N}{D}U\frac{d\hat{v}_{1x}}{dx}\right)+\frac{d}{dx}\left(\frac{W}{D}U \hat{v}_{1x}\right)
    +\left[\rho_0\omega^2+\rho_0'g+k_0^2\frac{\rho_0 g}{\rho_0\omega^2-k_0^2\frac{N}{D}}\left(\frac{W}{D}-\rho_0 g\right)-\left(\rho_0 gU\right)'\,\right]\hat{v}_{1x} = 0\,. \label{gen1Dvar}
\end{multline}
In this latter equation, we wrote 
\begin{equation}
    U= 1+ k_0^2 \frac{N/D}{\rho_0\omega^2-k_0^2\frac{N}{D}} = \frac{\rho_0\omega^2}{\rho_0\omega^2-k_0^2\frac{N}{D}}\,.
\end{equation}
We find the adiabatic limit through $N/D\rightarrow \gamma p_0$ while $W=0$, recovering textbook knowledge (eq.~(7.29) from \citet{goed2019}), and connecting to its recent extension to self-gravitating conditions \citep{Durrive2021}. Computing the spectrum now boils down to solving Eq.~(\ref{gen1Dvar}), which of course needs two boundary conditions, e.g. enforcing a vanishing velocity component perpendicular to the impenetrable walls at $x=0$ and $x=a$ by setting 
$\hat{v}_{1x}=0$ there, for a stratified medium in a slab of size $a$.

Interestingly, the factor $W$ prevents us to formulate the governing second order ODE for $\hat{v}_{1x}$ in the typical self-adjoint form $(p(x)y')'-q(x)y=0$. In fact, in published work on the thermal continuum in MHD settings, whether in plane-parallel atmospheres \citep{vdlslab1991} or in radially structured flux tubes \citep{vdl1991,vdl1991B,Hermans2024}, it was emphasized that we can manipulate the linearized MHD equations into this form. This is because all these works not only adopted a force-balanced equilibrium, but also a thermally balanced background. Indeed, if we specify to
\begin{equation}
    {\cal{L}}_0=0 \Rightarrow {\cal{L}}_0'= {\cal{L}}_\rho\rho_0'+{\cal{L}}_TT_0'=0 \Rightarrow W=i(1-\gamma) {\cal{L}}_T(T_0'\rho_0+\rho_0'T_0+\rho_0 g/{\cal{R}}) =0 \,,
\end{equation}
where the last equality invokes the force balance relation $p_0'=-\rho_0 g$. In this paper, we find that thermal misbalance in the background state leads to a more general governing second order ODE, with still the coefficient 
\begin{equation}
    \frac{N}{D}U=\frac{N\omega^2}{\omega^2D-k_0^2M} \label{coeff1}
\end{equation} in front of the highest order derivative. Note the difference between the denominator in Eq.~(\ref{coeff1}) (the factor $\omega^2 D-k_0^2 M$ containing $k_0^2$) with the uniform medium dispersion relation Eq.~(\ref{criteria}) (with $k^2$). This difference is crucially related to the presence of a direction of inhomogeneity, along which we can no longer use uncoupled, pure Fourier modes, but instead must compute the full eigenfunction variation. Thermal misbalance introduces a deviation from a Sturm-Liouville-type form that is proportionate to the factor $W$. We stress that the eigenvalue $\omega$ appears much more intricate in our governing equations than in the textbook Sturm-Liouville problem. The adiabatic case, as well as the case where we adopt exact thermal balance, does reduce to a self-adjoint type. 

For completeness, we also state the equivalent formulation in terms of two first order differential equations that link (derivates of) $\hat{p}_1$ and $\hat{v}_{1x}$. Indeed, we can write generally that
\begin{eqnarray}
    M\frac{d\hat{v}_{1x}}{dx} & = & \frac{\hat{v}_{1x}}{\rho_0}\left[-\omega\left(p_0'-c_0^2 \rho_0'\right)-M \rho_0'\right] +\frac{i \hat{p}_1}{\rho_0\omega}\left(\omega^2D-k_0^2 M\right) \,, \nonumber \\
    M\frac{d\hat{p}_{1}}{dx} & = & -g D \hat{p}_1+i\hat{v}_{1x}\left[ -g \left(p_0'-c_0^2 \rho_0'\right)+\omega\rho_0 M \right] \,. \label{2ode}
\end{eqnarray}
Starting from these, the same governing ODE Eq.~(\ref{gen1Dvar}) is obtained. This formulation makes it clear that a special role is played by the frequency range that we find from $M=0=N$, while nothing peculiar is at play for $D=0$. As in the more general cases discussed in our MHD textbook~\citep{goed2019} (in particular as noted on its page 476), the formulation~(\ref{2ode}) really only has one factor $M=0$ counting as a genuine continuum, since the determinant of the coefficients at the right hand side can be shown to be proportional to $M$.

\subsubsection{Special reference cases}
A special limit case could be to consider isothermal conditions, both in the equilibrium and in the perturbations. This is easily recovered by setting $\gamma=1$ (and $T_0$ constant), recognizing that then $T_1$ vanishes entirely and so does $S_1$, but $p_1/p_0=\rho_1/\rho_0$. This special case (which does not use the energy equation at all and sets a closure $p=c_i^2\rho$ with isothermal sound speed squared $c_i^2={\cal{R}}T_0$) is seen to reduce to the adiabatic regime, since $Q=0=W$ and $D=\omega$ whereas $N=p_0\omega$, and the governing equation~(\ref{gen1Dvar}) becomes simply
\begin{equation}
    \frac{d}{dx}\left(p_0\frac{\omega^2}{\omega^2-k_0^2c_i^2}\frac{d\hat{v}_{1x}}{dx}\right)
    +\left[\rho_0\omega^2+\rho_0'g-k_0^2\frac{\rho_0 g^2}{\omega^2-k_0^2c_i^2}-\left(\rho_0 g\right)' \frac{\omega^2}{\omega^2-k_0^2c_i^2}\,\right]\hat{v}_{1x}= 0\,.
\end{equation}
Hence, purely isothermal settings do not have thermal instability, as they do not involve the entropy law~(\ref{entr}) or (\ref{entrspec}).

Another well-known reference case, is the adiabatic case of an exponentially stratified atmosphere with constant sound speed (or temperature $T_0=constant$, but not assuming isothermality throughout), where $p_0=p_*e^{-\alpha x}$ and $\rho_0=\rho_*e^{-\alpha x}$, making $c_0^2=\gamma p_0/\rho_0=\gamma p_*/\rho_*$ constant. If also $g$ is constant then $\alpha\equiv\gamma g/c_0^2$ is constant. Then the governing equation~(\ref{gen1Dvar}) reduces to a second order ODE with constant coefficients \citep{goed2019}, written as
\begin{equation}
    \frac{d^2 \hat{v}_{1x}}{dx^2}-\alpha\frac{d\hat{v}_{1x}}{dx}+\frac{\omega^4-k_0^2 c_0^2\omega^2+k_0^2c_0^2N_B^2}{\omega^2 c_0^2}\hat{v}_{1x}=0\,.
\end{equation}
This introduced $N_B^2=(\gamma-1)g^2/c_0^2$ as constant and here always positive Brunt-V\"{a}is\"{a}l\"{a}a frequency, so that only stable waves remain. 
This has well known p-mode and g-mode solutions, and in a slab where $x\in[0,a]$ and with $\hat{v}_{1x}$ vanishing at $x=0$ and $x=a$ we get
\begin{equation}
    \omega^2_{p,g}=\frac{1}{2}c_0^2\left(k_0^2+\left(\frac{n\pi}{a}\right)^2+\frac{1}{4}\alpha^2\right)\left[1\pm \sqrt{1-\frac{4k_0^2N_B^2}{c_0^2\left(k_0^2+\left(\frac{n\pi}{a}\right)^2+\frac{1}{4}\alpha^2\right)^2}}\right]\,. \label{pg}
\end{equation}
The p-modes $\omega^2_p$ correspond to the plus sign for the term between brackets, while the g-modes take the minus sign. We note in particular that when $k_0=0$, g-modes are at marginal frequency $\omega=0$. 

\begin{figure}
    \centering
    \includegraphics[width=\linewidth]{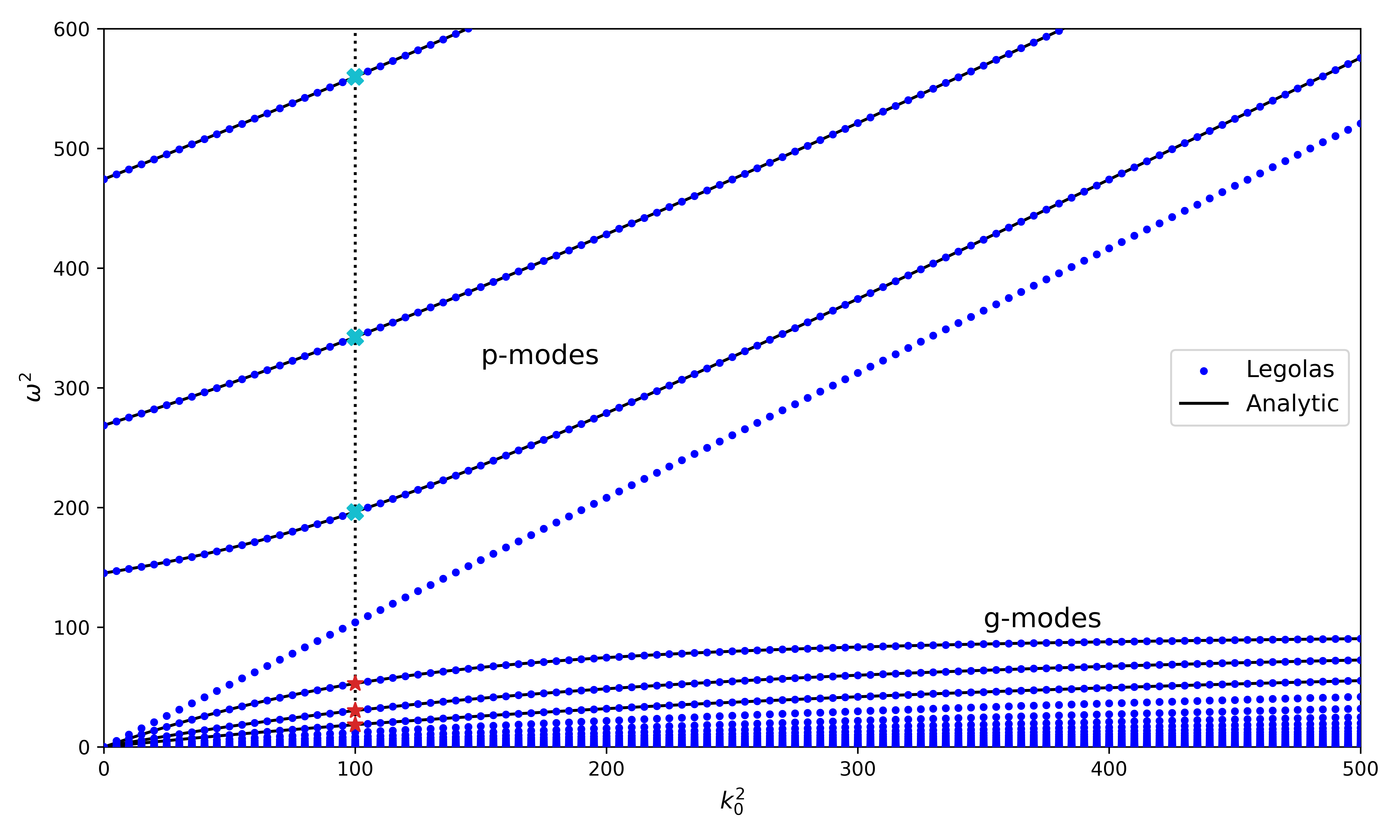}
    \caption{$\omega-k$ diagram for the exponential atmosphere presented as multiple \texttt{Legolas} runs. The analytic solutions of the p- and g-branches, given by Eq. (\ref{pg}), are shown for $n=1,2,3$. This analytic prediction for the first three p-mode branches, and the first three g-mode branches, is shown to agree with computed (blue dots) eigenfrequencies. The vertical, dotted line marks the spectrum shown in Fig. \ref{fig:exp-spectra}a. }
    \label{fig:pg-modes}
\end{figure}

\begin{figure}
    \centering
    \includegraphics[width=\linewidth]{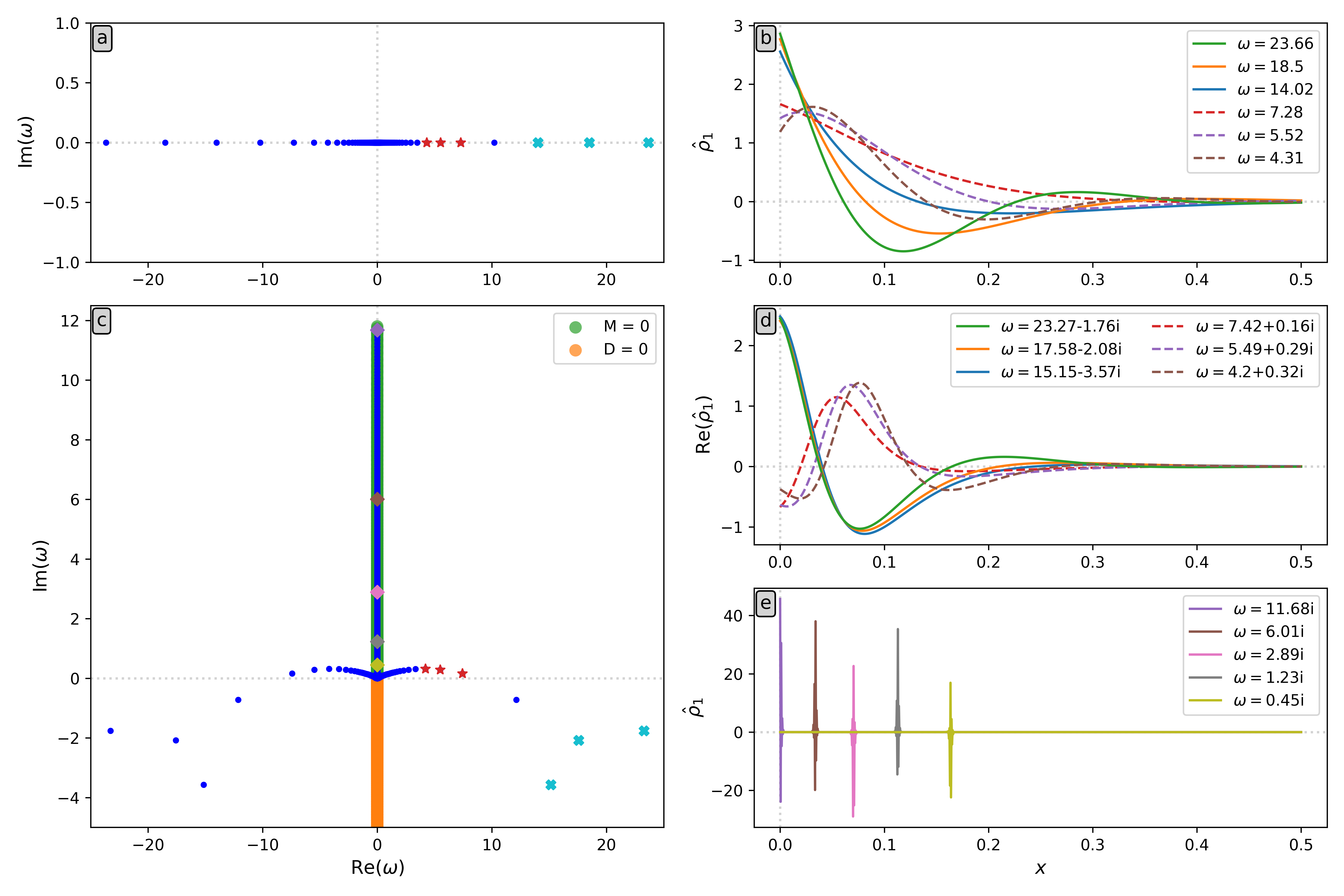}
    \caption{(a) Adiabatic $k_0^2 = 100$ spectrum of the exponentially stratified medium. Cyan crosses and red stars mark the forward-propagating $n=1,2,3$ p- and g-modes, respectively. (b) $\rho_1$ eigenfunctions of the p- (solid) and g-modes (dashed) marked in (a). (c) Modified $k_0^2 = 100$ spectrum including optically thin radiative cooling and constant background heating. Cyan crosses and red stars mark the modified, forward-propagating $n=1,2,3$ p- and g-modes, respectively. No modes lie in $D=0$ range (orange). (d) $\rho_1$ eigenfunctions of the damped p- (solid) and destabilized g-modes (dashed) marked in (c). (e) Singular $\rho_1$ eigenfunctions of a selection of $M=0$ continuum modes indicated as diamonds of the same color in (c).}
    \label{fig:exp-spectra}
\end{figure}

To see how the thermal continuum arises in non-adiabatic settings, we now present numerical spectra obtained from our general linear eigenvalue solver {\texttt{Legolas}} \citep{legolasA2020,legolasB}. For a (dimensionless) parameter choice of $\bar{\rho}_* = 10$, $\bar{p}_* = 6.25$, $\bar{\alpha} = 20$, and $\bar{g} = 12.5$, with normalizations $\rho_u = 4.82 \times 10^{-14}\ \mathrm{g}\,\mathrm{cm}^{-3}$, $T_u = 10^6\ \mathrm{K}$, and length $L_u = 10^9\ \mathrm{cm}$, for a gravitational acceleration of $g = 2.06\times 10^6\,\mathrm{cm}\,\mathrm{s}^{-2}$ ($\sim 100\,g_\mathrm{Sun}$), the dimensionless p- and g-mode diagram of the $\bar{x}\in [0, 0.5]$ slab is represented in Fig. \ref{fig:pg-modes} with $101$ \texttt{Legolas} runs with different $\bar{k}_0^2$ values, at $301$ grid points each. There is perfect agreement with the analytical results, as we overplotted the first three branches $n=1,2,3$ given by Eq.~(\ref{pg}). For the indicated wave number $\bar{k}_0^2 = 100$, the stable spectrum is shown in the complex eigenfrequency plane in Fig. \ref{fig:exp-spectra}a at $301$ grid points, showing the symmetry between left and rightward modes ($\pm \sqrt{\omega_{p,g}^2}$). The first three modes, $n=1,2,3$, of the p- and g-modes are marked with cyan crosses and red stars, respectively. Their corresponding $\rho_1$ eigenfunctions are given in Fig. \ref{fig:exp-spectra}b as solid (for p-modes) and dashed (g-modes) lines. When optically thin radiative cooling is included we no longer have the analytic results from Eq.~(\ref{pg}), and we must compute the entire eigenspectrum numerically, thereby choosing a particular heat-loss prescription ${\cal{L}}$. For optically thin radiative losses as appropriate for a solar or stellar corona, we use the \texttt{colgan\_dm} cooling curve combining the results of \citet{dalg1972} and \citet{colg2008}, as well as a constant heating balancing the cooling term such that $\mathcal{L}_0 = 0$. Doing so, we find that the spectrum becomes unstable, as shown in Fig. \ref{fig:exp-spectra}c at $501$ grid points. Here, we see that the p-modes, again marked with cyan crosses, are damped while the g-modes (red stars) become overstable. Secondly, the full $M=0$ range, shown in green, emerges as a thermal continuum along the imaginary axis in the unstable half-plane. The range $D=0$ is also indicated and is fully damped (it lies on the imaginary axis but in the lower half-plane), but it is important to note that this range does not contain any eigenmodes here. Similarly to the adiabatic case, Fig. \ref{fig:exp-spectra}d shows the real part of the $\rho_1$ eigenfunction of the selected p- (solid) and g-modes (dashed). Finally, the singular $\rho_1$ perturbations, indicative of continuum modes, of the thermal continuum are shown in Fig. \ref{fig:exp-spectra}e for the modes indicated in panel d with a diamond of the same color. The fastest growing continuum mode is singularly located near the bottom boundary, while each height gives a localized perturbation that grows exponentially.

\subsection{1D coronal `loops'}\label{sec:loop}

The general governing equation~(\ref{generalv1}) is also applicable to the 1D hydrodynamic ``coronal loop" case studied frequently in solar settings. There, the same set of governing equations is reduced to a triplet for density, pressure and projected velocity component, if we imagine that coronal magnetic loops are rigid and just set up pipe-flow along a coronal loop that may have its area varying along its length. We then can allow for the area variation $A(s)$ along the `field line' or `loop' given by an ad-hoc assumed field strength variation $B(s)$, and only have the projected gravity component $g_\parallel(s)$ prescribed according to the adopted field line shape. As the manipulations leading up to Eq.~(\ref{generalv1}) operated in vector form, we can directly write the governing second order ODE for the (only) velocity component as
\begin{equation}
    \partial_s\left[\frac{M}{D}\frac{1}{A}\partial_s\left(A\rho_0v_1\right)\right]+\partial_s\left[\frac{\omega v_1}{D}p_0\frac{\partial_s S_0}{S_0}\right]-g_{\parallel}\frac{1}{A}\partial_s\left(A\rho_0v_1\right)+\omega^2\rho_0v_1 =0 \,. \label{loopcase}
\end{equation}
This used the fact that divergence of a vector rewrites to $(1/A) \partial_s (A v_s)$ for a vector with only a $v_s$ component, while a gradient is just $\partial_s$.
We may note that this expression contains the entropy variation in the loop, which relates to the Brunt-V\"{a}is\"{a}l\"{a}a frequency through
$N_B^2(s)=-\frac{g_{\parallel}}{\gamma}\frac{\partial_sS_0}{S_0}$. It is a matter of straightforward algebra to further manipulate Eq.~(\ref{loopcase}) to the form that closely mimics Eq.~(\ref{gen1Dvar}) from the stratified atmosphere, namely
\begin{equation}
    \partial_s\left[\frac{N}{D}\partial_sv_1\right]+\partial_s\left[\left(\frac{W}{D}+\frac{N}{D}\frac{\partial_sA}{A}\right)v_1\right]+\left[\rho_0\omega^2+\rho_0\partial_sg_{\parallel}-\rho_0g_{\parallel}\frac{\partial_sA}{A}\right]v_1=0 \,. \label{loop2}
\end{equation}
We now see that Eq.~(\ref{gen1Dvar}) for the case where $k_0=0$ (where $U=1$) corresponds to this Eq.~(\ref{loop2}), when the area variation is ignored, i.e. $\partial_sA=0$. Now, both thermal misbalance (i.e. non-zero $W$) as well as non-trivial area variation imply that we cannot formulate a Sturm-Liouville-type equation. Still, the important factor in front of the highest order derivative vanishes when $M=0=N$, introducing the thermal continuum. As before, eigenmodes are obtained by solving Eq.~(\ref{loop2}) augmented with boundary conditions, which may assume line-tying where $v_1=0$ at both loop ends.

\begin{figure}
    \centering
    \includegraphics[width=1\linewidth]{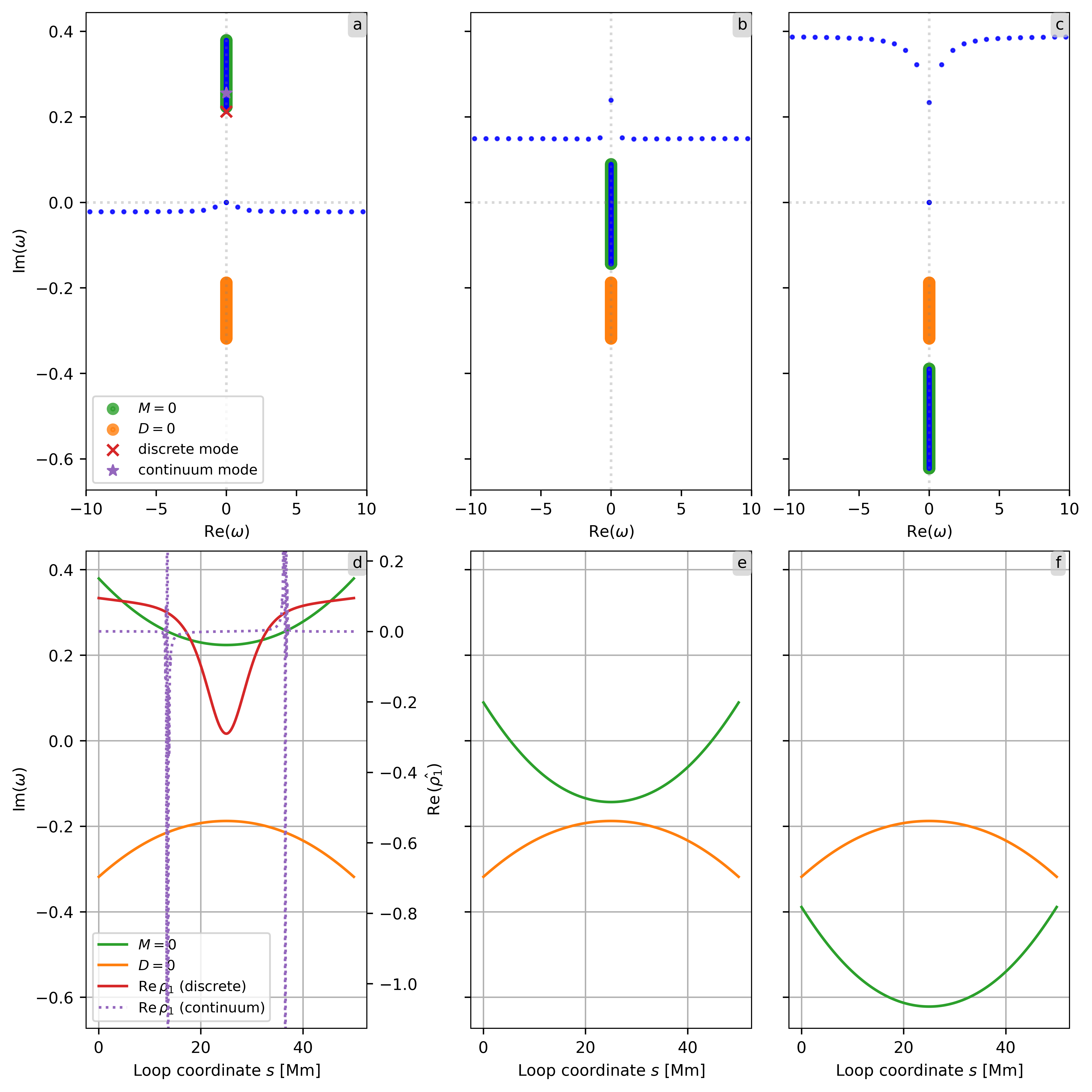}
    \caption{
(a) Thermal balance \(\mathcal{L}_0 = 0\).  The green region marks the unstable \(M=0\) thermal continuum and orange marks the damped \(D=0\) range. A red cross highlights the discrete thermal mode and a purple star a selected continuum mode. Blue dots indicate numerically computed eigenfrequencies.
(b) Over-heating with a uniform heating rate \(h_0 = 1.5\times10^{-3}\;\mathrm{erg\,cm^{-3}\,s^{-1}}\).  The entire \(M=0\) branch is shifted downward, whereas the \(D=0\) region remains unchanged. 
(c) Strong over-heating, \(h_0 = 3.0\times10^{-3}\;\mathrm{erg\,cm^{-3}\,s^{-1}}\), further shifts the thermal continuum into a damped regime. No eigenvalues appear within the \(D=0\) region in any case.
(d) Imaginary part of the continua for the balanced case, together with the real part of the density perturbation \(\hat{\rho_1}\) for the discrete (solid red) and continuum (dotted purple) modes highlighted in (a).  
(e) and (f) Same continua for the \(h_0 = 1.5\times10^{-3}\) and \(3.0\times10^{-3}\) cases. The shift of the \(M=0\) curve matches the displacement of the continuum in panels (b) and (c).}
    \label{fig:spectra-6panel}
\end{figure}

To quantify how non-adiabatic terms modify the MHD spectrum of a coronal loop, we again solve the linear MHD eigenvalue problem with {\texttt{Legolas}} \citep{legolasA2020,legolasB}.
In line with many previous nonlinear studies of condensation formation in coronal loops, we adopt a semicircular shape (inspired by solar coronal loop observations), but here only consider the coronal part of the loop. This allows to adopt a loop equilibrium with a constant temperature (in reality realized by effective thermal conduction along the field). This ignores the actual coupling to the denser chromosphere across a transition region. We also adopt a constant loop cross-section, and study spectra for line-tied loops, i.e. with vanishing velocity perturbations at both loop ends.
A semi-circular loop of physical length \(L = 50\;\mathrm{Mm}\) (radius \(a = L/\pi \simeq 15.9\;\mathrm{Mm}\)) is represented by a one-dimensional field-aligned coordinate \(s\in[0,L]\).
All quantities are nondimensionalized with the reference scales \(L_u = 10^{9}\,\mathrm{cm}\) ( \(= 10\ \mathrm{Mm}\) ), \(T_u = 10^{6}\ \mathrm{K}\) and \(n_u = 10^{9}\ \mathrm{cm^{-3}}\), with corresponding mass-density unit \(\rho_u = 1.67\times10^{-15}\,\mathrm{g\,cm^{-3}}\) and time unit \(t_u = L_u/\sqrt{k_B T_u/(\mu m_p)} \approx 110\) s. In these units the loop length and uniform background temperature become \(L = 5\) and \(T_{0}=1\) respectively, while the dimensionless gravity profile is \(g_\parallel(s)=g_\odot /(L_u/t_u^{2}) \cos (\pi s/L)\) with \(g_\odot = 274\ \mathrm{m\ s^{-2}}\).
The (dimensionless) equilibrium density \(\rho_0(s)\) is obtained by numerically integrating the isothermal force-balance equation  
\[
\frac{d p_0}{d s}=T_0\frac{d\rho_0}{ds}=\rho_0\,g_\parallel(s),
\]
ensuring a consistent hydrostatic background.
We next introduce optically thin losses through the \texttt{colgan\_dm} cooling curve and apply a heating rate \(h_0\).
Three heating scenarios are considered in Fig. \ref{fig:spectra-6panel}: (a) exact thermal balance, \(h_0=\rho_0\Lambda(T_0)\) so that \(\mathcal{L}_0=h_0-\rho_0\Lambda(T_0)=0\), (b) moderate constant heating, \(h_0 = 1.5\times10^{-3}\ \mathrm{erg\ cm^{-3}\ s^{-1}}\), and (c) strong heating, \(h_0 = 3.0\times10^{-3}\ \mathrm{erg\ cm^{-3}\ s^{-1}}\). Both latter choices consider cases with thermal imbalance where $\mathcal{L}_0\neq 0$. 
Along the loop, the radiative loss term spans \(\rho_{0}\Lambda(T_{0}) = (2.5\text{--}7.1)\times10^{-4}\,\mathrm{erg\,cm^{-3}\,s^{-1}}\), hence the adopted constant heating rates correspond to ratios \(h_{0}/[\rho_{0}\Lambda(T_{0})]\simeq 2.1\text{--}6.1\) and \(4.2\text{--}12.2\), respectively.
In the top panels (a,b,c), all blue dots are computed eigenfrequencies, of which there are always two types: damped (a) or overstable (b,c) {\em discrete p-modes}, as well as all numerically computed {\em continuum modes} (always found on the imaginary-$\omega$ axis). The latter correspond to the analytic prediction for the range $M=0$, while the `isochoric' range $D=0$ does not contain any (neither discrete, nor continuum) eigenmode for these examples.

We note from this figure that the effect of thermal misbalance is to modify the entire spectrum (both discrete and continuum modes). The change in the location of the continuum modes is as predicted by theory, influenced by the deviation from ${\cal{L}}_0=0$, while $D=0$ is not affected by misbalance, but also not relevant spectrally. 
Panels d-e-f show the corresponding variation of the frequency defined by the ranges $M=0$ and $D=0$ as function of the loop coordinate $s$. In panel d, we also show two eigenfunctions, one for a discrete mode found underneath the thermal continuum (in red), and one for a typical frequency in the continuum (in purple). The latter (purple) density eigenfunction shows clearly singular behavior at two locations along the loop (at $s\simeq 14$ and $s\simeq 36$ Mm). Since we placed this eigenfunction zero level at the height (in Im($\omega$)) of the selected continuum eigenfrequency, these locations are seen to match their intersection with the $M=0$ range (in green) exactly, in accord with analytic prediction for the localization of these singular modes.

\section{Including thermal conduction}\label{tcpart}
It is of interest to recall that \citet{Field1965} analyzed the uniform medium case, in both HD and MHD settings, where in the former isotropic, and in the latter anisotropic thermal conduction, were fully accounted for. The original work that discovered the thermal continuum in non-uniform MHD cases equally accounted for anisotropic thermal conduction effects \citep{vdl1991,vdl1991B,vdlslab1991}. Starting from the nonlinear governing temperature evolution equation~(\ref{Tevol}), we have in hydro
\begin{eqnarray}
    {\cal{R}}\rho \partial_t T +{\cal{R}}\rho\bma{v}\cdot\nabla T + (\gamma-1)p\nabla\cdot \bma{v} &=& \left(\gamma-1\right)\left[\rho{\cal{L}}+\nabla\cdot\left(\kappa^c\nabla T\right)\right] \label{Tevol2}\,,
\end{eqnarray}
where we could adopt an isotropic thermal conductivity that depends on temperature such as $\kappa^c(T)\propto (T/T_u)^{2.5}$. In the whole linearization procedure, we now need to add the linearized conduction term. When we denote $\frac{d\kappa^c}{dT}\equiv\kappa^c_T$ and $\frac{d^2\kappa^c}{dT^2}\equiv\kappa^c_{TT}$ and use the zero subscript to indicate its evaluation in the background state, the generalization of Eq.~(\ref{idT}) becomes
\begin{multline}
      \left[i D+(\gamma-1)\frac{\kappa^c_{T0}\nabla^2T_0+\kappa^c_{TT0}|\nabla T_0|^2}{{\cal{R}}\rho_0}\right] \frac{T_1}{T_0} +(\gamma-1)\frac{2\kappa^c_{T0}\nabla T_0\cdot\nabla T_1+\kappa^c_0\nabla^2 T_1}{p_0} \\  
      \hspace*{3cm}=  \bma{v}_1 \cdot \frac{\nabla S_0}{S_0} + \left[iD(\gamma-1)-\frac{Q}{p_0}\right]\frac{\rho_1}{\rho_0} \,. 
\end{multline}
The terms on the left hand side containing spatial gradients of $T_1$ complicate the manipulation of the linearized equations, which will no longer be reducible to a governing second order ODE for a perturbed velocity. It is likely that a spatially fourth order version exists, but it becomes obvious that the numerical approach as used in {\texttt{Legolas}} is now preferable: we do not expect more normal modes to appear, but rather a modification (likely a damping) of the corresponding modes without conduction included. 

In uniform setting where we can introduce a wavevector and its wavenumber $k$, it is easily shown that the dispersion relation~(\ref{druni}) modifies to 
\begin{equation}
    \omega^3-i\omega^2\frac{1}{C_v}\left({\cal{L}}_T-\frac{k^2\kappa^c_0}{\rho_0}\right)-\omega c_0^2 k^2-i(\gamma-1)\left({\cal{L}}_0+\rho_0{\cal{L}}_\rho-T_0\left({\cal{L}}_T-\frac{k^2\kappa^c_0}{\rho_0}\right)\right)k^2=0 \,, \label{drunikappa}
\end{equation}
as already presented in \citet{Field1965}. The change in spatial order is evident since we go from a second to a fourth order polynomial in terms of the wavenumber $k$. In \citet{Begelman1990}, the authors introduced the so-called Field length, which is directly found from dimensional analysis on the right-hand side of Eq.~(\ref{Tevol2}), and is given by
\begin{equation}
    \lambda_F=\sqrt{\frac{\kappa(T) T}{\rho {\cal{L}}}} \,.
\end{equation}
Note that \citet{Field1965} never used this particular lengthscale, which only makes sense for ${\cal{L}}\ne0$, and it was therefore replaced by using the maximal (and strictly positive) net heating or cooling rate in its denominator. It quantifies a lengthscale that determines whether thermal conduction, or rather external heating and radiative cooling (i.e. ${\cal{L}}$) is dominant for the temperature structure. In many works that analyze the role of thermal instability from the uniform medium dispersion relation as in Eq.~(\ref{drunikappa}), one then distinguishes if one has the wavenumber $k$ such that its lengthscale is (much) smaller or larger than this Field length \citep{Waters2019}. \cite{Falle2020} illustrated how a more relevant critical wavelength expression (also appearing in \citet{Field1965,Begelman1990}) can be obtained using Whitham theory of wave hierarchies, from which the various known (isochoric, isobaric or isentropic) instability criteria are clearly identifiable.

We stress that when computing eigenmodes using {\texttt{Legolas}}, the inclusion of thermal conduction is possible by implementing and activating the corresponding terms in the set of linearized equations to solve, and ensuring that the boundary conditions are consistent with the physics included. In the weak Galerkin formalism using finite element discretizations for all eigenfunctions, we thus far only handled essential boundary conditions, allowing to set e.g. specific velocity components to zero (realizing line-tying). Specific for the 1D hydro loop model, activating thermal conduction implies the need to generalize the pure-line-tied boundary conditions in {\texttt{Legolas}} to allow for additional Neumann boundary conditions on the gradient of the temperature perturbation, expected to vanish in the presence of thermal conduction, and analyse its effect on the non-adiabatic spectrum for 1D stratified loops. Such natural boundary conditions still need to be implemented.

This issue does not arise in cylindrical flux tubes with only radially varying magnetic fields, as they do not have a radial magnetic field component (which crosses the tube boundary) in equilibrium. Hence, earlier studies for the MHD case in~\citet{vdl1991,vdl1991B,vdlslab1991} could show that including finite perpendicular thermal conduction removes the thermal continuum, and replaces it with a dense set of discrete modes that show intricate fine-structure across the magnetic surfaces: a natural explanation for the fine structure in prominences. 

It is to be expected that inclusion of thermal conduction in the 1D loop settings yields a similar result, where instead of a true continuum range, a very dense set of eigenmodes with still extreme localization and rapid variation is obtained. This is left for future work. This type of analysis should then also allow for the entire chromosphere to corona background temperature variation of the loop, and not just address the isothermal coronal part as done here. We note that this likely requires extreme resolution through the transition region, as all eigenmodes and eigenfunctions will be fully aware of all relevant lengthscales in the background state.

\section{Discussion and future extensions}\label{sec:lform}
\subsection{The role of heat-loss functions}
Except for the {\texttt{Legolas}}-based results, we never really quantified the precise terms in the net heating-cooling function ${\cal{L}}$. This is done deliberately, to highlight that all analytical findings above are completely general, and could be extended easily to handle effects we ignored thus far. We recall that we assumed a fixed equation of state for an ideal gas, while varying ionization and composition effects may well be relevant depending on the situation at hand (galactic, interstellar, stellar or solar coronal). Furthermore, we assumed that the net heating-cooling function is in essence only a function of ${\cal{L}}(\rho,T)$. In the context of solar coronal loops, we often encounter $\rho{\cal{L}}=H-n_e n_H\Lambda(T)$ for an assumed heating function $H$ (per unit volume, or $H=\rho h$ with $h$ per unit mass). This includes radiative losses due to optically thin radiation as pre-computable in a tabulated loss function $\Lambda(T)$. In the literature, many such cooling curves $\Lambda(T)$ have been adopted, with \citet{herm2021} providing an incomplete list of twenty such curves. The plasma composition and ionization will determine the relation between density $\rho$ and the electron $n_e$ and proton $n_H$ number densities, so that varying composition and ionization fractions also play a role here. Countless nonlinear numerical studies in 1D loops have focused on the precise spatial dependence of the external heating function $H$, and it is known that footpoint localization and its location with respect to the coronal transition region (where thermal conduction sets downwards heat fluxes into the chromosphere) matter for coronal condensation formation~\citep{Antolin2020,Pelouze2022}.  It would be of interest to make similar parametric surveys of how the entire spectrum (including the thermal continuum) changes as a result of varying heating prescriptions.

\subsection{The role of thermal misbalance}
The incorporation of thermal misbalance (i.e. the consequences of ${\cal{L}}_0\ne0$ and corresponding equation~(\ref{T0evol})) allows many possibilities, with timescales $\tau_0=C_vT_0/{\cal{L}}_0$ of the background state in competition with instantaneous growthrates found in the spectrum. This implies that a perfect force and energy balance is not within reach, although this is usually adopted in traditional spectral linear stability analysis (but we relaxed this explicitly in this work and in \cite{Keppens2016}). In the literature on coronal loop condensations (rain and prominences), various authors instead describe the resulting (non-linear) time-dependent evolution  as a state of `thermal non equilibrium' or TNE \citep{Antiochos2000,Klimchuk2019}. However, the interplay between a spatio-temporally changing background and a correspondingly changing eigenmode spectrum, are hardly ever explored, and will likely provide a deeper understanding of how the linear stability reacts to a nonlinearly changing background. This contrasts with attempts to differentiate between TNE and thermal instability \citep{Klimchuk2019}. Limit cycles exist and are found in many simulations. We believe that these can be understood from advanced spectroscopic analysis, since the thermal continuum is at all times a robust (and quantifiable in growthrate) ingredient of the linear eigenmode spectrum, and will play a role in the obtained nonlinear evolutions. We note in passing that observed scaling laws in coronal loops suggest coronal loop heating functions with dependencies $H(\rho,T,B)\propto (\rho/\rho_u)^a (T/T_u)^b (B/B_u)^c$ as quantified by power-law indices $(a,b,c)$ (see e.g.~\citet{Kolotkov2023,Brughmans2022}). While these can be adopted in analysis of the dispersion relations for uniform media, in non-uniform stratified settings, adopting such heating directly almost surely implies thermal misbalance (i.e. no local match with $n_en_H\Lambda(T)$ making ${\cal{L}}\ne0$) in much of the atmosphere. Therefore, nonlinear numerical simulations targeting coronal rain or prominence formation usually relax initial conditions first to achieve an almost thermal balance throughout the entire atmosphere using some balancing background heating, to then trigger condensations by thermal instability in a controlled manner through a more localized or impulsive heating prescription \citep{Jercic2024,li2022}.

\subsection{Astrophysically relevant extensions}
As further extensions, we note that our external gravitational field $\bma{g}$ may also incorporate an acceleration due to a radiation force, such as appropriate for massive star stellar winds \citep{Moens2022}. In this paper, we assumed a purely spatial dependence $\bma{g}(\bma{x})$, so the effective reduction of the gravitational field by a constant factor $(1-\Gamma_e)$ due to a high stellar luminosity and associated Eddington parameter $\Gamma_e=\kappa_e L_*/4\pi GM_*c$ (with fundamental constants of gravity $G$ and speed of light $c$) for a star with luminosity $L_*$ and mass $M_*$ and a (constant) electron scattering opacity $\kappa_e$  is -in principle- included. For massive star outflows, incorporating line-driven effects would introduce another effective gravity term with dependencies of the form $\bma{g}_{line}(\nabla\bma{v},\rho)$, a case that can be studied in future work. 

 Extensions to radiative hydro settings where the form of the heat-loss function ${\cal{L}}$ was generalized to allow for varying opacity effects were explored in \cite{Proga2022}, to allow for free-free cooling and Compton heating.
Similarly, we may consider extensions to a two-temperature model for plasma-radiation energy exchange such as adopted in a flux limited diffusion approximation \citep{MoensFLD2022}. Then we have ${\cal{L}}=c a_r(\kappa_E T_r^4-\kappa_P T^4)$ involving the energy $\kappa_E$ and Planck $\kappa_P$ mean opacities, where the radiation field energy density is expressed as $a_rT_r^4$ using a radiation temperature $T_r$ (as opposed to the gas temperature $T$). In those expressions $a_r$ is the radiation constant and $c$ the speed of light. Therefore, the function ${\cal{L}}$ would depend on three thermodynamic variables ${\cal{L}}(\rho, T, T_r)$, and the detailed thermodynamic dependencies of the opacities bring in many possibilities for further radiatively driven instabilities.

\section{Summary and Outlook}

In this paper, we presented a rigorous analysis of the governing spectral equations for normal modes in stratified, non-adiabatic hydrodynamic settings. We confronted analytical findings that identified the thermal continuum, with numerical results for gravitationally stratified atmospheres, and for coronal loop models. Our results serve to emphasize the importance of the spectral (normal mode) view on loop or atmosphere stability, as influenced by radiative losses.

It will be of interest to relate these spectral insights more firmly to fully non-linear thermal non-equilibrium evolutions in coronal loop settings \citep{Antolin2022,Jercic2024} or in stellar atmospheres \citep{Simon2024}. This linear-to-nonlinear link has been initiated in recent work where combined thermal and resistive tearing instabilities interplay in magnetized current sheets in resistive, non-adiabatic MHD \citep{Jonghe2025}. Our purely hydrodynamic findings also help to understand the dominance of thermal instability in realistically stratified chromosphere-to-corona solar magnetized atmospheres \citep{Claes2021}, implying that thermal instability may well explain its intricate multi-thermal aspect. The results here also connect to studies emphasizing the role of thermal misbalance in thin-flux-tube guided wave descriptions \citep{Kolotkov2023}, which again involve mostly hydrodynamic equations.

Combining detailed linear spectral analysis with fully nonlinear evolutions of 1D solar coronal loop models, by quantifying spectra at consecutive instances of nonlinear simulations, and by contrasting it with purely linear time evolutions, should also clarify the expected, intricate link between thermal non-equilibrium and thermal instability, although some authors argue that both are ``fundamentally different" \citep{Klimchuk2019}.  
The role of isochoric versus isobaric evolutions and what may cause the thermal instability to saturate in hydrodynamic settings has been discussed recently by \citet{Waters2023}, with reasoning mostly based on the dispersion relation for a homogeneous, non-adiabatic gas, combined with the nonlinear entropy evolution equation. The spectral complexities of non-homogeneous background states, such as highlighted here, are yet to be appreciated. It is in that respect intriguing that we find that in actually stratified settings, there is no true isochoric continuum (i.e. the entire $D=0$ range does not form part of the eigenspectrum) but the thermal continuum we locate as $M=0$ mostly relates to isobaric conditions.
Our findings dismiss the role of the ``isochoric catastrophic cooling" mode \citep{Waters2025}, as it does not form part of the normal mode spectrum, except in the unrealistic infinite medium setting where it appears as a spurious mode that does not vanish at infinity. Curiously, a number of studies in solar settings have meanwhile confirmed that the onset of spontaneous condensation formation is well described by the thermal instability as approximated by the isochoric criterion from \cite{Field1965}: in 1D hydrodynamic coronal loops by \citet{Xia2011}, in 2D prominences forming in magnetic arcades \citep{xia2012}, or in fully sub-photosphere to coronal radiative MHD simulations of active region coronal rain by \citet{zekun2024}. Analyzing the obtained nonlinear simulations using phase-space views, such as done recently for flux-rope embedded prominences forming by levitation-condensation by \citet{Brughmans2022}, should help to understand how both isochoric and isobaric conditions may play a role in the full nonlinear evolutions and how the (time-evolving) eigenmode spectrum relates to this.

 As far as extending this work to flowing equilibria, i.e. a state with $\bma{v}_0\ne \bma{0}$, \citet{Balbus1986} pointed out that the relevant criterion to consider for local thermal instability is the variation of $\delta({\cal{L}}/T)$, and this was then applied to a spherical outflow regime. This is clear from Eq.~(\ref{entrspec}). Follow-up work can make direct contact with those findings, where we then need to incorporate the role of background flow to the area-expanding hydro loop settings, or study how non-adiabatic effects alter stability of transonic flows like the solar wind. Our hydrodynamic treatment did not yet include background flow effects. However, recent studies of linear stability in magnetized cylinders or loops do include flow as well as non-adiabatic effects \citep{Hermans2024}, showing that in such (assuming thermal balance) case, the thermal continuum gets Doppler shifted into the complex eigenfrequency plane. 

 Other meaningful extensions of this work are to consider the role played by spatially-varying partial ionization effects, which may in first instance be incorporated through ambipolar diffusion mimicking ion-neutral interactions, as in \citet{Ballester2024} where thermal misbalance was studied for propagating MHD waves. For coronal loops that span from the low chromosphere to high into the solar corona, this effect is clearly important. Related to this, the assumption of purely optically thin radiative losses should be generalized as well, and in that respect findings on local radiative hydro and MHD instabilities in optically thick conditions \citep{Blaes2003} are to be connected with the optically thin scenarios adopted here or in many solar coronal applications.

\section*{Acknowledgment} 
       RK acknowledges funding from the KU Leuven C1 project C16/24/010 UnderRadioSun and the Research Foundation Flanders FWO project G0B9923N Helioskill. JDJ is supported by Research Foundation - Flanders (FWO) fellowship 1225625N. NB is supported by Research Foundation - Flanders (FWO) fellowship 11J2624N. We thanks the referee for a careful reading and constructive feedback.

\bibliography{TCmain_final}{}

\begin{thebibliography}{}
\expandafter\ifx\csname natexlab\endcsname\relax\def\natexlab#1{#1}\fi
\providecommand{\url}[1]{\href{#1}{#1}}
\providecommand{\dodoi}[1]{doi:~\href{http://doi.org/#1}{\nolinkurl{#1}}}
\providecommand{\doeprint}[1]{\href{http://ascl.net/#1}{\nolinkurl{http://ascl.net/#1}}}
\providecommand{\doarXiv}[1]{\href{https://arxiv.org/abs/#1}{\nolinkurl{https://arxiv.org/abs/#1}}}

\bibitem[{{Antiochos} {et~al.}(2000){Antiochos}, {MacNeice}, \&
  {Spicer}}]{Antiochos2000}
{Antiochos}, S.~K., {MacNeice}, P.~J., \& {Spicer}, D.~S. 2000, \apj, 536, 494,
  \dodoi{10.1086/308922}

\bibitem[{{Antolin}(2020)}]{Antolin2020}
{Antolin}, P. 2020, Plasma Physics and Controlled Fusion, 62, 014016,
  \dodoi{10.1088/1361-6587/ab5406}

\bibitem[{{Antolin} {et~al.}(2022){Antolin}, {Mart{\'\i}nez-Sykora}, \&
  {{\c{S}}ahin}}]{Antolin2022}
{Antolin}, P., {Mart{\'\i}nez-Sykora}, J., \& {{\c{S}}ahin}, S. 2022, \apjl,
  926, L29, \dodoi{10.3847/2041-8213/ac51dd}

\bibitem[{{Appert} {et~al.}(1974){Appert}, {Gruber}, \&
  {Vaclavik}}]{Appert1974}
{Appert}, K., {Gruber}, R., \& {Vaclavik}, J. 1974, Physics of Fluids, 17,
  1471, \dodoi{10.1063/1.1694918}

\bibitem[{{Balbus}(1986)}]{Balbus1986}
{Balbus}, S.~A. 1986, \apjl, 303, L79, \dodoi{10.1086/184657}

\bibitem[{{Ballester} {et~al.}(2024){Ballester}, {Soler}, {Terradas}, \&
  {Carbonell}}]{Ballester2024}
{Ballester}, J.~L., {Soler}, R., {Terradas}, J., \& {Carbonell}, M. 2024,
  Philosophical Transactions of the Royal Society of London Series A, 382,
  20230222, \dodoi{10.1098/rsta.2023.0222}

\bibitem[{{Begelman} \& {McKee}(1990)}]{Begelman1990}
{Begelman}, M.~C., \& {McKee}, C.~F. 1990, \apj, 358, 375,
  \dodoi{10.1086/168994}

\bibitem[{{Blaes} \& {Socrates}(2003)}]{Blaes2003}
{Blaes}, O., \& {Socrates}, A. 2003, \apj, 596, 509, \dodoi{10.1086/377637}

\bibitem[{{Brughmans} {et~al.}(2022){Brughmans}, {Jenkins}, \&
  {Keppens}}]{Brughmans2022}
{Brughmans}, N., {Jenkins}, J.~M., \& {Keppens}, R. 2022, \aap, 668, A47,
  \dodoi{10.1051/0004-6361/202244071}

\bibitem[{{Case}(1960)}]{Case1960}
{Case}, K.~M. 1960, Physics of Fluids, 3, 143, \dodoi{10.1063/1.1706010}

\bibitem[{{Claes} {et~al.}(2020{\natexlab{a}}){Claes}, {De Jonghe}, \&
  {Keppens}}]{legolasA2020}
{Claes}, N., {De Jonghe}, J., \& {Keppens}, R. 2020{\natexlab{a}}, \apjs, 251,
  25, \dodoi{10.3847/1538-4365/abc5c4}

\bibitem[{{Claes} \& {Keppens}(2019)}]{Claes2019}
{Claes}, N., \& {Keppens}, R. 2019, \aap, 624, A96,
  \dodoi{10.1051/0004-6361/201834699}

\bibitem[{{Claes} \& {Keppens}(2021)}]{Claes2021}
---. 2021, \solphys, 296, 143, \dodoi{10.1007/s11207-021-01894-2}

\bibitem[{{Claes} \& {Keppens}(2023)}]{legolasB}
---. 2023, Computer Physics Communications, 291, 108856,
  \dodoi{10.1016/j.cpc.2023.108856}

\bibitem[{{Claes} {et~al.}(2020{\natexlab{b}}){Claes}, {Keppens}, \&
  {Xia}}]{Claes2020}
{Claes}, N., {Keppens}, R., \& {Xia}, C. 2020{\natexlab{b}}, \aap, 636, A112,
  \dodoi{10.1051/0004-6361/202037616}

\bibitem[{{Colgan} {et~al.}(2008){Colgan}, {Abdallah}, {Sherrill}, {Foster},
  {Fontes}, \& {Feldman}}]{colg2008}
{Colgan}, J., {Abdallah}, J., J., {Sherrill}, M.~E., {et~al.} 2008, \apj, 689,
  585, \dodoi{10.1086/592561}

\bibitem[{{Daley-Yates} \& {Jardine}(2024)}]{Simon2024}
{Daley-Yates}, S., \& {Jardine}, M.~M. 2024, \mnras, 534, 621,
  \dodoi{10.1093/mnras/stae2131}

\bibitem[{{Dalgarno} \& {McCray}(1972)}]{dalg1972}
{Dalgarno}, A., \& {McCray}, R.~A. 1972, \araa, 10, 375,
  \dodoi{10.1146/annurev.aa.10.090172.002111}

\bibitem[{{De Jonghe} \& {Sen}(2025)}]{Jonghe2025}
{De Jonghe}, J., \& {Sen}, S. 2025, \mnras, 536, 3308,
  \dodoi{10.1093/mnras/stae2740}

\bibitem[{{Donn{\'e}} \& {Keppens}(2024)}]{donn2024}
{Donn{\'e}}, D., \& {Keppens}, R. 2024, \apj, 971, 90,
  \dodoi{10.3847/1538-4357/ad50a3}

\bibitem[{{Durrive} {et~al.}(2021){Durrive}, {Keppens}, \&
  {Langer}}]{Durrive2021}
{Durrive}, J.-B., {Keppens}, R., \& {Langer}, M. 2021, \mnras, 506, 2336,
  \dodoi{10.1093/mnras/stab1726}

\bibitem[{{Falle} {et~al.}(2020){Falle}, {Wareing}, \& {Pittard}}]{Falle2020}
{Falle}, S.~A.~E.~G., {Wareing}, C.~J., \& {Pittard}, J.~M. 2020, \mnras, 492,
  4484, \dodoi{10.1093/mnras/staa131}

\bibitem[{{Field}(1965)}]{Field1965}
{Field}, G.~B. 1965, \apj, 142, 531, \dodoi{10.1086/148317}

\bibitem[{{Goedbloed} {et~al.}(2019){Goedbloed}, {Keppens}, \&
  {Poedts}}]{goed2019}
{Goedbloed}, H., {Keppens}, R., \& {Poedts}, S. 2019, {Magnetohydrodynamics of
  Laboratory and Astrophysical Plasmas} (Cambridge University Press),
  \dodoi{10.1017/9781316403679}

\bibitem[{{Goedbloed}(1998)}]{Hans1998}
{Goedbloed}, J.~P. 1998, Physics of Plasmas, 5, 3143, \dodoi{10.1063/1.873041}

\bibitem[{{Hermans} \& {Keppens}(2021)}]{herm2021}
{Hermans}, J., \& {Keppens}, R. 2021, \aap, 655, A36,
  \dodoi{10.1051/0004-6361/202140665}

\bibitem[{{Hermans} \& {Keppens}(2024)}]{Hermans2024}
---. 2024, \aap, 686, A180, \dodoi{10.1051/0004-6361/202348337}

\bibitem[{{Jer{\v{c}}i{\'c}} {et~al.}(2024){Jer{\v{c}}i{\'c}}, {Jenkins}, \&
  {Keppens}}]{Jercic2024}
{Jer{\v{c}}i{\'c}}, V., {Jenkins}, J.~M., \& {Keppens}, R. 2024, \aap, 688,
  A145, \dodoi{10.1051/0004-6361/202348442}

\bibitem[{{Keppens} \& {Demaerel}(2016)}]{Keppens2016}
{Keppens}, R., \& {Demaerel}, T. 2016, Physics of Plasmas, 23, 122117,
  \dodoi{10.1063/1.4971811}

\bibitem[{{Klimchuk}(2019)}]{Klimchuk2019}
{Klimchuk}, J.~A. 2019, \solphys, 294, 173, \dodoi{10.1007/s11207-019-1562-z}

\bibitem[{{Kolotkov} {et~al.}(2023){Kolotkov}, {Nakariakov}, \&
  {Fihosy}}]{Kolotkov2023}
{Kolotkov}, D.~Y., {Nakariakov}, V.~M., \& {Fihosy}, J.~B. 2023, Physics, 5,
  193, \dodoi{10.3390/physics5010015}

\bibitem[{{Li} {et~al.}(2022){Li}, {Keppens}, \& {Zhou}}]{li2022}
{Li}, X., {Keppens}, R., \& {Zhou}, Y. 2022, \apj, 926, 216,
  \dodoi{10.3847/1538-4357/ac41cd}

\bibitem[{{Lu} {et~al.}(2024){Lu}, {Chen}, {Guo}, {Ding}, {Wang}, {Yu}, {Ni},
  \& {Xia}}]{zekun2024}
{Lu}, Z., {Chen}, F., {Guo}, J.~H., {et~al.} 2024, \apjl, 973, L1,
  \dodoi{10.3847/2041-8213/ad73d2}

\bibitem[{{Miki{\'c}} {et~al.}(2013){Miki{\'c}}, {Lionello}, {Mok}, {Linker},
  \& {Winebarger}}]{Mikic2013}
{Miki{\'c}}, Z., {Lionello}, R., {Mok}, Y., {Linker}, J.~A., \& {Winebarger},
  A.~R. 2013, \apj, 773, 94, \dodoi{10.1088/0004-637X/773/2/94}

\bibitem[{{Moens} {et~al.}(2022{\natexlab{a}}){Moens}, {Poniatowski},
  {Hennicker}, {Sundqvist}, {El Mellah}, \& {Kee}}]{Moens2022}
{Moens}, N., {Poniatowski}, L.~G., {Hennicker}, L., {et~al.}
  2022{\natexlab{a}}, \aap, 665, A42, \dodoi{10.1051/0004-6361/202243451}

\bibitem[{{Moens} {et~al.}(2022{\natexlab{b}}){Moens}, {Sundqvist}, {El
  Mellah}, {Poniatowski}, {Teunissen}, \& {Keppens}}]{MoensFLD2022}
{Moens}, N., {Sundqvist}, J.~O., {El Mellah}, I., {et~al.} 2022{\natexlab{b}},
  \aap, 657, A81, \dodoi{10.1051/0004-6361/202141023}

\bibitem[{{Parker}(1953)}]{Parker1953}
{Parker}, E.~N. 1953, \apj, 117, 431, \dodoi{10.1086/145707}

\bibitem[{{Pelouze} {et~al.}(2022){Pelouze}, {Auch{\`e}re}, {Bocchialini},
  {Froment}, {Miki{\'c}}, {Soubri{\'e}}, \& {Voyeux}}]{Pelouze2022}
{Pelouze}, G., {Auch{\`e}re}, F., {Bocchialini}, K., {et~al.} 2022, \aap, 658,
  A71, \dodoi{10.1051/0004-6361/202140477}

\bibitem[{{Peng} \& {Matsumoto}(2017)}]{Peng2017}
{Peng}, C.-H., \& {Matsumoto}, R. 2017, \apj, 836, 149,
  \dodoi{10.3847/1538-4357/aa5be8}

\bibitem[{{Proga} {et~al.}(2022){Proga}, {Waters}, {Dyda}, \&
  {Zhu}}]{Proga2022}
{Proga}, D., {Waters}, T., {Dyda}, S., \& {Zhu}, Z. 2022, \apjl, 935, L37,
  \dodoi{10.3847/2041-8213/ac87b0}

\bibitem[{{Tan} {et~al.}(2023){Tan}, {Oh}, \& {Gronke}}]{Gronke2023}
{Tan}, B., {Oh}, S.~P., \& {Gronke}, M. 2023, \mnras, 520, 2571,
  \dodoi{10.1093/mnras/stad236}

\bibitem[{{van der Linden} \& {Goossens}(1991{\natexlab{a}})}]{vdl1991B}
{van der Linden}, R.~A.~M., \& {Goossens}, M. 1991{\natexlab{a}}, \solphys,
  134, 247, \dodoi{10.1007/BF00152647}

\bibitem[{{van der Linden} \& {Goossens}(1991{\natexlab{b}})}]{vdlslab1991}
---. 1991{\natexlab{b}}, \solphys, 131, 79, \dodoi{10.1007/BF00151746}

\bibitem[{{van der Linden} {et~al.}(1991){van der Linden}, {Goossens}, \&
  {Goedbloed}}]{vdl1991}
{van der Linden}, R.~A.~M., {Goossens}, M., \& {Goedbloed}, J.~P. 1991, Physics
  of Fluids B, 3, 866, \dodoi{10.1063/1.859842}

\bibitem[{{Wareing} {et~al.}(2017){Wareing}, {Pittard}, \&
  {Falle}}]{Wareing2017}
{Wareing}, C.~J., {Pittard}, J.~M., \& {Falle}, S.~A.~E.~G. 2017, \mnras, 470,
  2283, \dodoi{10.1093/mnras/stx1417}

\bibitem[{{Waters} \& {Proga}(2019)}]{Waters2019}
{Waters}, T., \& {Proga}, D. 2019, \apj, 875, 158,
  \dodoi{10.3847/1538-4357/ab10e1}

\bibitem[{{Waters} \& {Proga}(2023)}]{Waters2023}
---. 2023, Frontiers in Astronomy and Space Sciences, 10, 1198135,
  \dodoi{10.3389/fspas.2023.1198135}

\bibitem[{{Waters} \& {Stricklan}(2025)}]{Waters2025}
{Waters}, T., \& {Stricklan}, A. 2025, \solphys, 300, 5,
  \dodoi{10.1007/s11207-024-02417-5}

\bibitem[{{Wibking} {et~al.}(2025){Wibking}, {Voit}, \& {O'Shea}}]{Wibking2025}
{Wibking}, B.~D., {Voit}, G.~M., \& {O'Shea}, B.~W. 2025, \mnras, 537, 739,
  \dodoi{10.1093/mnras/staf092}

\bibitem[{{Xia} {et~al.}(2012){Xia}, {Chen}, \& {Keppens}}]{xia2012}
{Xia}, C., {Chen}, P.~F., \& {Keppens}, R. 2012, \apjl, 748, L26,
  \dodoi{10.1088/2041-8205/748/2/L26}

\bibitem[{{Xia} {et~al.}(2011){Xia}, {Chen}, {Keppens}, \& {van
  Marle}}]{Xia2011}
{Xia}, C., {Chen}, P.~F., {Keppens}, R., \& {van Marle}, A.~J. 2011, \apj, 737,
  27, \dodoi{10.1088/0004-637X/737/1/27}

\end{thebibliography}
\bibliographystyle{aasjournal}

\end{document}